\documentclass[12pt]{article}
\usepackage{amsfonts}
\usepackage{epsfig}
\begin{document}
\title{\bf Exact solution\\ to the Landau-Lifshitz equation\\
in a constant electromagnetic field}
\author{Yurij Yaremko\footnote{Electronic mail: yar@icmp.lviv.ua}}
\date{Institute for Condensed Matter Physics\\ Svientsitskii Str. 1, 79011 Lviv, Ukraine}
\maketitle
\begin{abstract}
We are interested in the motion of a classical charge acted upon an external constant electromagnetic field where the back
reaction of the particle's own field is taken into account. The Landau-Lifshitz approximation to the Lorentz-Abraham-Dirac
equation is solved exactly and in closed form. It is shown that the ultrarelativistic limit of the Landau-Lifshitz equation
for a radiating charge is the equation for eigenvalues and eigenvectors of the external electromagnetic field tensor.
\end{abstract}
\section{Introduction}\label{intro}

There has been considerable interest in the classical motion of a radiating charge in strong electromagnetic field which is
stimulated by various astrophysical situations such as stellar jets, cosmic rays, radiation from pulsars, etc. The other region
of applicability is the behavior of charged particles in man-made devices, such as Penning traps and accelerators. There has
been considerable interest in dynamics of a charged particle exposed to an ultraintense laser
field \cite{HL10,KS98}. The radiation damping plays dominant role in ultrarelativistic regime.

The most prominent and widely accepted equation of motion of a point charged particle acted upon an external force as well as
its own electromagnetic field is the Lorentz-Abraham-Dirac (LAD) equation \cite{Dir}:
\begin{equation}\label{LAD}
a^\mu=\frac{1}{m}f^\mu_{\rm ext}+\tau_0\left[\dot{a}^\mu-(a\cdot a)u^\mu\right].
\end{equation}
The factor by which the Abraham radiation reaction 4-vector is scaled,
\begin{equation}
\tau_0=\frac{2e^2}{3mc^3},
\end{equation}
is a constant with dimension of time, whose numerical value for the electron is $\tau_0=6.24\cdot 10^{-24} s$. Because of the
third time derivative, the equation is beyond the Newtonian class of equations of motions which completely specify the trajectory of a particle once the initial position and velocity are given. LAD possesses pathological solutions \cite{Rohr,Parr,Vill02}, such as runaway solution (when acceleration grows exponentially with time) and preacceleration (when acceleration begins to increase prior the time at which the external force switches on).

Using the successive iteration, Landau and Lifshitz \cite[\S 76]{LL}
(LL) replace the square of acceleration by the negative scalar
product $(u\cdot\dot{a})$ and substitute $m^{-1}{\rm d}f^\mu_{\rm
ext}/{\rm d}\tau$ for the derivative of particle's acceleration:
\begin{equation}\label{LL-eq}
a^\mu=\frac{1}{m}f^\mu_{\rm ext}+\tau_0\left(\delta^\mu{}_\nu+u^\mu u_\nu
\right)\frac{1}{m}\frac{{\rm d}f^\nu_{\rm ext}}{{\rm d}\tau}.
\end{equation}
The LL equation is of Newtonian class which avoids the nonphysical solutions of the LAD \cite{Sp00} (see also Refs. \cite{Rh01}
and \cite{Spohn}). The rigorous derivation of the complete first-order correction to the Lorentz force equation in the external
field was recently given \cite{GHW09} using perturbation theory. For a structureless point particle, Eq. (\ref{LL-eq}) has been
obtained.

There are several analytic solutions to the Landau-Lifshitz equation. Ares de Parga and Mares \cite{Ares} give the solution to
the LL equation for the radiating charge in static and spatially uniform electromagnetic field. Rivera and Villarroel \cite{RV02}
find out the solutions to LL equation for two circling charges confined by specific electromagnetic field at the opposite ends of
diameter. Rajeev \cite{Raj} solves the non-relativistic approximation of the LL equation for a charged particle moving in a
Coulomb potential. Di Piazza \cite{DiP} gives the exact solution to the LL equation for a charge acted upon by a plane wave of
arbitrary shape and polarization. Similar problem has been considered by Hadad et al. in Ref. \cite{HL10} devoting to theoretical
studies of radiation emission from free electrons in very intense laser pulses.

Ares de Parga and Mares \cite{Ares} present the solution to
Eq.(\ref{LL-eq}) in the form $u^\mu(\tau)=\eta(\tau)f^\mu(\tau)$
where $u^\mu$ are components of particle's 4-velocity, scalar
function $\eta(\tau)$ represents the damping effect, and 4-vector
$f^\mu(\tau)=\exp(\alpha(\tau))f^\mu_L(\tau)$ is proportional to a
solution $f^\mu_L(\tau)$ of the Lorentz force equation. Because
exponential factor, the expression differs from Chen's solution
\cite[eq.(11)]{Shen} to the ``truncated'' LL equation
\cite[eqs.(9),(10)]{Shen} where the intermediate term is omitted.

In this paper, we study the Ares de Parga and Mares solution
\cite[eq. (12)]{Ares}. In the present paper we use the
Heaviside-Lorentz system of units and the mostly plus Minkowski
metrics $g_{\alpha\beta}={\rm diag}(-1,1,1,1)$, and the equation
looks as follows
\begin{equation}\label{APM}
u^\mu(\tau)=\frac{f^\mu(\tau)}{\sqrt{1-2\tau_0\frac{e^2}{m^2}\int_0^\tau{\rm d}s f_\nu(s)F^\nu{}_\alpha F^\alpha{}_\beta
f^\beta(s)}}\,,
\end{equation}
where
\begin{equation}\label{ff0}
f^\mu(\tau)=\exp\left[\left(\frac{e}{m}F^\mu{}_\alpha+\tau_0\frac{e^2}{m^2}F^\mu{}_\nu F^\nu{}_\alpha\right)
\tau\right]f^\alpha(0)\,.
\end{equation}
The exponential function of the combination of the linear and squared applied field can be presented as series in powers of
the electromagnetic field tensor $F$. To sum up the series we follow the algebraic approach sketched in Ref. \cite{Shen}.
According to the Hamilton-Cayley theorem, the Faraday tensor (\ref{FldT}) satisfies its own characteristic equation
\begin{equation}\label{chrc}
F^4+{\cal S}F^2-\frac14{\cal P}^2I=0,
\end{equation}
where ${\cal P}$ and ${\cal S}$ are invariants of this tensor given by eqs. (\ref{SP}). Hence, any power of $F$ may be reduced
to a linear combination of the unit matrix $I$, $F$ itself, its square $F^2$, and its cube $F^3$. The goal of this paper is
to present the formal solution (\ref{APM}) as a linear combination of $n$-tuples of $F$ no higher than cubic.

The paper is organized as follows. In Section \ref{Fradkin}, we review the basic algebraic properties of the electromagnetic
field tensor such as its eigenvalues and eigenvectors. Following Ref. \cite{Frad}, we construct two projection operators which
allow us to split the Minkowski space into two sets of mutually orthogonal planes. (The plane $(x^0x^3)$ and its orthogonal
$(x^1x^2)$ offer a good illustration of this splitting.) In Section \ref{motion}, we apply the projection technique for solving
of the LL equation for a radiating charge in a static and spatially uniform electromagnetic field. We split the equation into
two equations described the orbits in planes mentioned above. The equations are completely uncoupled to each other and we solve
them separately. In Section \ref{Applicat}, we study three main examples: motion in the field of magnetic type, in the field of
electric type, and in the crossed field. In Section \ref{Concl}, we summarize the main ideas and results.

\section{Fradkin's projection operators}\label{Fradkin}
\setcounter{equation}{0}
At each point of Minkowski space ${\mathbb M}_{\,4}$, the electromagnetic field is determined by the Faraday 2-form $\hat F$
and its Hodge dual 2-form $\!\!\!\phantom{1}^{\,*}\!{\hat F}$. If the Minkowski rectangular coordinates are adapted, the state
of electromagnetic field ${\hat F}=\frac12F_{\alpha\beta}\,{\rm d}x^\alpha\wedge{\rm d}x^\beta$ at point $x\in{\mathbb M}_{\,4}$ is specified by three components $(E^1,E^2,E^3)$ of electric field and three components $(B^1,B^2,B^3)$ of magnetic field
\begin{equation}\label{Far}
(F_{\alpha\beta})=\left(
\begin{array}{cccc}
0&-E^1&-E^2&-E^3\\
E^1&0&B^3&-B^2\\
E^2&-B^3&0&B^1\\
E^3&B^2&-B^1&0\\
\end{array}
\right).
\end{equation}
Its Hodge dual $\!\!\!\phantom{1}^{\,*}\!{\hat F}=\frac12{\hat\omega}({\mathbf F})$ is 2-form, whose components are
\begin{equation}\label{Fst}
(^{\,*\!}F_{\mu\nu})=\left(
\begin{array}{cccc}
0&B^1&B^2&B^3\\
-B^1&0&E^3&-E^2\\
-B^2&-E^3&0&E^1\\
-B^3&E^2&-E^1&0\\
\end{array}
\right).
\end{equation}
Before we introduce the projection operators providing by the electromagnetic field tensor (\ref{Far}), we recall its basic algebraic properties.

\subsection{Invariants}\label{Inv}

The matrices (\ref{Far}) and (\ref{Fst}) describe not only the electromagnetic field. They bear the imprint of the inertial frame which is used to determine the components of electric and magnetic fields. There exist two scalar functions which do not depend on the choice of basis. To construct the {\it invariants} of the electromagnetic field, we apply the Hodge star operator to 4-forms ${\hat F}\wedge{\hat F}$ and ${\hat F}\wedge\!\!\!\phantom{1}^*\!{\hat F}$. The manipulation results in two 0-forms
\begin{eqnarray}
{\cal P}&=&\!\!\!\phantom{1}^*\!({\hat F}\wedge{\hat F})\nonumber\\
&=&\frac12F_{\mu\nu}\!\!\!\phantom{1}^*\!F^{\mu\nu},\label{P}\\
{\cal S}&=&\!\!\!\phantom{1}^*\!({\hat F}\wedge\!\!\!\phantom{1}^*\!{\hat F})\nonumber\\
&=&\frac12F_{\mu\nu}F^{\mu\nu}.\label{S}
\end{eqnarray}
For practical considerations, it is simple and convenient to express ${\cal P}$ and ${\cal S}$ in terms of three-vectors
$\mathbf{E}$ and $\mathbf{B}$:
\begin{equation}\label{SP}
{\cal P}=2\,(\mathbf{E}\,\mathbf{B}),\qquad\qquad {\cal S}={\mathbf{B}}^2-{\mathbf{E}}^2.
\end{equation}

What are the consequences of the fact that the invariants ``overlook'' any transformations of basis? Electromagnetic fields
are divided into classes according to whether ${\cal P}=0$ or ${\cal P}\ne 0$ as well as whether ${\mathbf{E}}^2$ or
${\mathbf{B}}^2$ is greater. If the electric field and the magnetic field are mutually orthogonal in a given inertial frame,
they are orthogonal in any other frame of reference. Indeed, despite a change of basis caused by a Poincar\'e transformation,
the value of scalar product of three-vectors $\mathbf{E}$ and $\mathbf{B}$ remains zero. The other invariant is the difference
between squared magnitudes of magnetic and electric fields. So, if they are equal to each other in a given Lorentz frame, they
are equal in any other frame of reference. The commonly used and widely accepted classification is as follows:

\begin{itemize}
\item[(A)] ${\cal P}=0,\quad {\cal S}\ne 0.$

Since the sign of ${\cal S}$ is invariant, there are two possibilities:
\begin{itemize}
\item[(a)]
If ${\cal S}<0$, then in all reference frames the magnitude of electric field is greater than the magnitude of magnetic field. The field is said to be of {\it electric type}.
\item[(b)]
If ${\cal S}>0$, then in all reference frames the magnitude of magnetic field is greater than that of electric field. The field is said to be of {\it magnetic type}.
\end{itemize}
\item[(B)] ${\cal P}\ne 0,\quad {\cal S}\ne 0.$ Such a field belongs to the set of {\it crossed} fields.
\item[(C)] ${\cal P}=0,\quad {\cal S}=0.$

If both the invariants ${\cal S}$ and ${\cal P}$  are equal to zero, we deal with the {\it null} field.
\end{itemize}

\subsection{Eigenvalues and eigenvectors}\label{EE}

Eigenvalues and eigenvectors of the electromagnetic field tensor are crucially important parts of the projection technique
\cite{Frad}. The eigenvalues of ${1\choose 1}$ tensor $F^\mu{}_\beta=g^{\mu\alpha}F_{\alpha\beta}$ are the roots of the
fourth-degree polynomial:
\begin{equation}\label{eig_v}
\eta^4+{\cal S}\eta^2 -\frac14 {\cal P}^2=0,
\end{equation}
where ${\cal S}$ and ${\cal P}$ are invariants (\ref{SP}). The set of eigenvalues consists of two pairs, one real and the other
imaginary
\begin{equation}\label{n_eig}
\{\eta\}=\{+b,-b,+{\rm i} a,-{\rm i} a\},
\end{equation}
where $a$ and $b$ are the scalar functions of invariants:
\begin{eqnarray}\label{aeig}
a&=&\sqrt{\frac12\left(S+\sqrt{{\cal S}^2+{\cal P}^2}\right)},\\
b&=&\sqrt{\frac12\left(-S+\sqrt{{\cal S}^2+{\cal P}^2}\right)}.\label{beig}
\end{eqnarray}
If ${\cal P}=0$, the set (\ref{n_eig}) contains doubly degenerate zero eigenvalue
\begin{itemize}
\item
for a field of electric type (${\cal S}< 0$) : $\{\eta\}=\{+\sqrt{-{\cal S}},-\sqrt{-{\cal S}},0,0\}$;
\item
for a field of magnetic type (${\cal S}> 0$) : $\{\eta\}=\{0,0,+{\rm i} \sqrt{\cal S},-{\rm i} \sqrt{\cal S}\}$.
\end{itemize}
A null electromagnetic field (both the invariants ${\cal P}=0$ and ${\cal S}=0$) is characterized by the quadruply
degenerate zero eigenvalue.

The eigenvector of the square matrix
\begin{equation}\label{FldT}
(F^\alpha{}_\nu)=\left(
\begin{array}{cccc}
0&E^1&E^2&E^3\\
E^1&0&B^3&-B^2\\
E^2&-B^3&0&B^1\\
E^3&B^2&-B^1&0
\end{array}
\right),
\end{equation}
is the non-zero vector, say $g$, that after being transformed by the matrix remains parallel to the original vector
\begin{equation}\label{eig_e}
F^\mu{}_\alpha g^\alpha=\eta_{(o)} g^\mu.
\end{equation}
The factor $\eta_{(o)}$ by which the eigenvector is scaled is a
particular element of the eigenvalue set (\ref{n_eig}). If $g$ is an
eigenvector of the field tensor with eigenvalue $\eta_{(o)}$, then
the multiplication of $g$ on any number is also an eigenvector of
$F$ with the same eigenvalue. All the eigenvectors of the Faraday
tensor are four-vectors of null lengths.

\subsubsection{Projection operators}\label{Proj}

According to the Hamilton-Cayley theorem, the Faraday tensor (\ref{FldT}) satisfies characteristic equation (\ref{chrc}).
This matrix polynomial can be presented as the product of four factors \cite[eq. (25)]{Frad}
\begin{equation}\label{chrm}
\left(F-\eta_{(o)}I\right)\left(F+\eta_{(o)}I\right)\left(F-\frac{{\rm i} ab}{\eta_{(o)}}I\right)\left(F+\frac{{\rm i} ab}{\eta_{(o)}}I\right)=0,
\end{equation}
where $\eta_{(o)}$ is a particular element of the eigenvalue set
(\ref{n_eig}). The equation suggests the form of the projection
operator producing eigenvector with a given eigenvalue $\eta_{(o)}$.
Since the eigenvector satisfies the equation $F-\eta_{(o)}I=0\,,$
the desired matrix operator should be proportional to the product of
the remaining matrices \cite[eq. 26]{Shen}
\begin{equation}
P_{(o)}=\prod_{\kappa\neq o}^4\frac{F-\eta_{(\kappa)}I}{\eta_{(o)}-\eta_{(\kappa)}}.
\end{equation}

Fradkin \cite{Frad} rewrites the characteristic equation (\ref{chrm}) in the form
\begin{equation}\label{FbFa}
\left(F^2-b^2I\right)\left(F^2+a^2I\right)=0,
\end{equation}
and fabricates two operators
\begin{eqnarray}
O^{(a)}&=&-\frac{F^2-b^2I}{a^2+b^2},\label{Oa}\\
O^{(b)}&=&\frac{F^2+a^2I}{a^2+b^2},\label{Ob}
\end{eqnarray}
which satisfy the projection algebra
\begin{eqnarray}
O^{(a)}O^{(a)}&=&O^{(a)},\nonumber\\
O^{(b)}O^{(b)}&=&O^{(b)},\nonumber\\
O^{(a)}+O^{(b)}&=&I,\label{AOa}
\end{eqnarray}
and
\begin{eqnarray}\label{AOab}
O^{(a)}O^{(b)}&=&O^{(b)}O^{(a)}\nonumber\\
&=&0.
\end{eqnarray}
The operator $O^{(a)}$ when acting on an arbitrary four-vector, say $g$, produces the four-vector
$$
g_{(a)}=O^{(a)}g,
$$
which is a linear combination of eigenvectors associated with the
imaginary eigenvalues $+{\rm i} a$ and $-{\rm i} a$. Indeed, the
operator
\begin{equation}\label{Papl}
P_{(+a)}=\frac{F+{\rm i} aI}{2{\rm i} a}O^{(a)},
\end{equation}
transforms the 4-vector $g$ into eigenvector $g_{(+a)}$ with eigenvalue $+{\rm i} a$ while
\begin{equation}\label{Pamn}
P_{(-a)}=-\frac{F-{\rm i} aI}{2{\rm i} a}O^{(a)},
\end{equation}
produces eigenvector $g_{(-a)}$  with negative imaginary eigenvalue
$-{\rm i} a$. Since $P_{(+a)}+P_{(-a)}=O^{(a)}$, the projective
operator (\ref{Oa}) separates two-dimensional subspace ${\cal
M}_x^{(a)}\subset{\rm T}_x{\mathbb M}_{\,4}$ which is spanned by
``imaginary'' eigenvectors $g_{(+a)}$ and $g_{(-a)}$.

Analogously, $O^{(b)}$ builds two-dimensional subspace ${\cal
M}_x^{(b)}\subset{\rm T}_x{\mathbb M}_{\,4}$ spanned by ``real''
eigenvectors $g_{(+b)}=P_{(+b)}g$ and $g_{(-b)}=P_{(-b)}g$ where
\begin{equation}\label{Pbmn}
P_{(+b)}=\frac{F+bI}{2b}O^{(b)},\qquad P_{(-b)}=-\frac{F-bI}{2b}O^{(b)}.
\end{equation}
It is obviously that their sum is the projective operator (\ref{Ob}).

Since eq. (\ref{AOab}), the four-vectors $g_{(a)}=O^{(a)}g$ and $g_{(b)}=O^{(b)}g$ are orthogonal to each other:
$$
(g_{(a)}\cdot g_{(b)})=0.
$$
At each point $x\in{\mathbb M}_{\,4}$ a vector from the tangent
space ${\rm T}_x{\mathbb M}_{\,4}$ can be presented as a sum of
vectors from ${\cal M}_x^{(a)}$ and ${\cal M}_x^{(b)}$.

\section{Charge in static uniform electromagnetic field}\label{motion}
\setcounter{equation}{0} In this Section, we apply the Fradkin's
operators (\ref{Oa}) and (\ref{Ob}) to the Landau-Lifshitz equation
for a radiating charge in a constant electromagnetic field. If the
field strengths which constitute the tensor (\ref{FldT}) are  static
and spatially uniform, its eigenvectors do not change with points of
Minkowski space. The subspaces spanned by these eigenvectors
constitute foliation of ${\mathbb M}_{\,4}$ by two-dimensional
planes. Traveling from point to point, we construct the coordinate
grid from these planes which covers all the flat spacetime. The
particle's world line may be decomposed into two orbits in these
mutually orthogonal two-dimensional sheets.

The Landau-Lifshitz equation where the proper time derivative of the
4-acceleration is replaced by the derivative of the contraction of
tensor (\ref{FldT}) and particle's 4-velocity
\begin{eqnarray}
\frac{1}{m}\frac{{\rm d}f^\mu_{\rm ext}}{{\rm d}\tau}&=&\frac{e}{m}F^\mu{}_\alpha a^\alpha\nonumber\\
&=&\frac{e^2}{m^2}F^\mu{}_\alpha F^\alpha{}_\beta u^\beta.\label{F-F}
\end{eqnarray}
is the so-called Herrera equation \cite[eq. (4)]{Ares}
\begin{equation}\label{LLeq}
\dot{u}^\mu=\frac{e}{m}F^\mu{}_\alpha u^\alpha+\tau_0\left(\delta^\mu{}_\nu+u^\mu u_\nu
\right)\frac{e^2}{m^2}F^\nu{}_\alpha F^\alpha{}_\beta u^\beta .
\end{equation}
At first, we decompose the particle's four-velocity $u$ into two mutually orthogonal four-vectors
\begin{equation}\label{uuab}
u^\mu_{(a)}=O^{(a)}u^\mu\qquad {\rm and} \qquad u^\mu_{(b)}=O^{(b)}u^\mu\,.
\end{equation}
Since $u=u_{(a)}+u_{(b)}$ and $(u\cdot u)=-1$, then
$u^2_{(a)}+u^2_{(b)}=-1$. The  projection operators (\ref{Oa}) and
(\ref{Ob}) commute with differential operator ${\rm d}/{\rm d}\tau$
(for the constant field only) as well as with the field tensor $F$
(for any field). To manipulate with the product $F\cdot F$, we use
the relations
\begin{eqnarray}\label{F2O}
F^2O^{(a)}&=&O^{(a)}F^2\qquad {\rm and} \qquad F^2O^{(b)}=O^{(b)}F^2\nonumber\\
&=&-a^2O^{(a)},\qquad\qquad\qquad\qquad =b^2O^{(b)},
\end{eqnarray}
which can be readily derived from the characteristic equation
(\ref{FbFa}). So, acted on the expression (\ref{F-F}) by operators
$O^{(a)}$ and $O^{(b)}$, one after another, we obtain
\begin{eqnarray}
O^{(a)}\frac{1}{m}\frac{{\rm d}f^\mu_{\rm ext}}{{\rm d}\tau}&=&-\omega_a^2 u^\mu_{(a)}\,,\nonumber\\
O^{(b)}\frac{1}{m}\frac{{\rm d}f^\mu_{\rm ext}}{{\rm d}\tau}&=&\lambda_b^2 u^\mu_{(b)}\,,\nonumber
\end{eqnarray}
where
$$
\omega_a=\frac{e}{m}a,\qquad \lambda_b=\frac{e}{m}b,
$$
and constants $a$ and $b$ are given by the expressions (\ref{aeig}) and (\ref{beig}), respectively.
It is easy to show that contraction of 4-vector (\ref{F-F}) with particle's 4-velocity is as follows:
\begin{equation}
\frac{1}{m}\frac{{\rm d}f^\mu_{\rm ext}}{{\rm d}\tau}u_\mu=-\omega_a^2 u^2_{(a)}+\lambda_b^2 u^2_{(b)}\,,
\end{equation}
where $u^2_{(a)}$ and $u^2_{(b)}$ are the squared norms of 4-velocity projections.

Inserting these in Eq. (\ref{LLeq}), we split it into two
second-order differential equations
\begin{eqnarray}
\frac{{\rm d} u^\mu_{(a)}}{{\rm d}\tau}&=&\frac{e}{m}F^\mu{}_\alpha u^\alpha_{(a)}-\omega_0\left[1+u^2_{(a)}\right]u^\mu_{(a)}\,,\label{LLua}\\
\frac{{\rm d} u^\mu_{(b)}}{{\rm d}\tau}&=&\frac{e}{m}F^\mu{}_\alpha u^\alpha_{(b)}+\omega_0\left[1+u^2_{(b)}\right]u^\mu_{(b)}\,,\label{LLub}
\end{eqnarray}
where constant
\begin{equation}\label{Om_0}
\omega_0=\tau_0\left(\omega_a^2+\lambda_b^2\right)
\end{equation}
determines the intensity of the radiation damping. We take into account that $u^2_{(a)}+u^2_{(b)}=-1$.

The two projections of Eq. (\ref{LLeq}) are completely uncoupled
with each other and we solve them separately. Following Eq.~(11) in
Ref.~\cite{Shen}, we express the solutions in the form
\begin{equation}\label{uab}
u^\mu_{(a)}(\tau)=\sqrt{A(\tau)}f^\mu_{(a)}(\tau)\,, \qquad u^\mu_{(b)}(\tau)=\sqrt{-B(\tau)}f^\mu_{(b)}(\tau)\,,
\end{equation}
where 4-vectors $f_{(a)}$ and $f_{(b)}$ are normalized to $+1$ and
$-1$, respectively. Inserting $u_{(a)}$ and $u_{(b)}$ in Eqs.
(\ref{LLua}) and (\ref{LLub}), respectively, we derive the simple
equations on the factors $A$ and $B$
\begin{eqnarray}\label{ABLfL}
\frac12\frac{{\rm d} A}{{\rm d}\tau}&=&-\omega_0\left[1+A\right]A,\nonumber\\
\frac12\frac{{\rm d} B}{{\rm d}\tau}&=&\omega_0\left[1+B\right]B.
\end{eqnarray}
These ordinary differential equations are supplemented with initial conditions
\begin{eqnarray}
A(0)&=&u^2_{(a)}(0)\qquad\qquad B(0)=u^2_{(b)}(0)\nonumber\\
&=&\alpha, \qquad\qquad\qquad\qquad\,\, =-\beta.\nonumber
\end{eqnarray}
The positive constants $\alpha$ and $\beta$ are related by the condition $\beta=\alpha+1$.

Equations (\ref{ABLfL}) can be solved in a simple way and we do not
go into detail. Contrary to the unperturbed Lorentz-force equation,
the squared norms $u^2_{(a)}:=A$ and $u^2_{(b)}:=B$ of four-velocity
projections change with time
\begin{eqnarray}\label{ABua}
u_{(a)}^2(\tau)&=&\frac{\alpha}{\beta{\rm e}^{2\omega_0\tau}-\alpha},\\
u_{(b)}^2(\tau)&=&-\frac{\beta}{\beta-\alpha{\rm e}^{-2\omega_0\tau}}.\label{ABub}
\end{eqnarray}
It is worth noting that the authors \cite[Eqs.(18)-(21)]{Ares}
derive the same expressions in Applications where the pure magnetic
field and pure electric field have been considered.

Inserting Eqs. (\ref{uab}) into Eqs. (\ref{LLua}) and (\ref{LLub}),
we see that the 4-vectors $f^\mu_{(a)}$ and $f^\mu_{(b)}$ satisfy
the unperturbed Lorentz-force equation
$$
\frac{{\rm d}f_{(a)}^\mu}{{\rm d}\tau}=\frac{e}{m}F^\mu{}_\alpha f_{(a)}^\alpha\,,\qquad
\frac{{\rm d}f_{(b)}^\mu}{{\rm d}\tau}=\frac{e}{m}F^\mu{}_\alpha f_{(b)}^\alpha\,.
$$
The sum of the two equations is the equation on the total 4-vector $f=f_{(a)}+f_{(b)}$
$$
\frac{{\rm d}f^\mu}{{\rm d}\tau}=\frac{e}{m}F^\mu{}_\alpha f^\alpha.
$$
In comparison to the Ares de Parga and Mares solution (\ref{ff0}),
the solution to this equation simplifies
\begin{equation}\label{ff0L}
f^\mu(\tau)=\exp\left(\frac{e}{m}F^\mu{}_\alpha\tau\right)f^\alpha(0).
\end{equation}
The reason of the clear distinction is that the particle's
4-velocity is the sum of two projections of 4-vector $f(\tau)$,
$f_{(a)}(\tau)$, and $f_{(b)}(\tau)$, scaled by two quite different
factors, $\sqrt{A}$ and $\sqrt{-B}$, (see Eqs. (\ref{uab})). While
in the the Ares de Parga and Mares scheme the 4-velocity is supposed
to be proportional to $f$ itself.

Adapting the Rosen's result \cite[eq.(1.8)]{Ros} for a Hermitian
matrix to the Faraday matrix (\ref{FldT}), Shen \cite[Appendix
B]{Shen} presents the action of exponential matrix operator as the
combination of a complete set of eigenvalues (\ref{n_eig}) and its
appropriate projection operators, one for each eigenvalue
\begin{eqnarray}
f(\tau)&=&\exp\left(\frac{e}{m}F\tau\right)f(0)\nonumber\\
&=&\sum_{o=1}^4\exp\left(\frac{e}{m}\eta_{(o)}\tau\right)P_{(o)}f(0).
\end{eqnarray}
According to eqs. (\ref{uab}), the ``a''-projection of particle's
4-velocity is proportional to the ``a''-projection of auxiliary
4-vector $f(\tau)$
\begin{eqnarray}
u_{(a)}(\tau)&=&\sqrt{A}f_{(a)}(\tau)\nonumber\\
&=&\sqrt{A}\left({\rm e}^{{\rm i}\omega_a\tau}P_{(+a)}+{\rm e}^{-{\rm i}\omega_a\tau}P_{(-a)}\right)f(0),\nonumber
\end{eqnarray}
while $u_{(b)}$ is proportional to $f_{(b)}$:
\begin{eqnarray}
u_{(b)}(\tau)&=&\sqrt{-B}f_{(b)}(\tau)\nonumber\\
&=&\sqrt{-B}\left({\rm e}^{\lambda_b\tau}P_{(+b)}+{\rm e}^{-\lambda_b\tau}P_{(-b)}\right)f(0).\nonumber
\end{eqnarray}
Because of $\left.A(\tau)\right|_{\tau=0}=\alpha$ and
$\left.B(\tau)\right|_{\tau=0}=-\beta$, the initial values
$f_{(a)}^\alpha(0)=u_{(a)}^\alpha(0)/\sqrt{\alpha}$ and
$f_{(b)}^\alpha(0)=u_{(b)}^\alpha(0)/\sqrt{\beta}$.

Substituting the right-hand sides of eqs. (\ref{Papl})-(\ref{Pbmn})
for the projection operators $P_{(o)}$, we finally obtain the
solutions to the Landau--Lifshitz equation (\ref{LLeq}):
\begin{eqnarray}
u^\mu_{(a)}(\tau)&=&\frac{1}{\sqrt{\beta{\rm e}^{2\omega_0\tau}-\alpha}}\left[u^\mu_{(a)}(0)\cos(\omega_a\tau)+a^{-1}F^\mu{}_\alpha u^\alpha_{(a)}(0)\sin(\omega_a\tau)\right],\nonumber\\
u^\mu_{(b)}(\tau)&=&\frac{1}{\sqrt{\beta-\alpha{\rm e}^{-2\omega_0\tau}}}\left[u^\mu_{(b)}(0)\cosh(\lambda_b\tau)+b^{-1}F^\mu{}_\alpha u^\alpha_{(b)}(0)\sinh(\lambda_b\tau)\right].\nonumber\\\label{uObL}
\end{eqnarray}
The radiation reaction free solutions in between the square brackets
coincide with the Fradkin's solutions \cite[eqs.(3.8),(3.9)]{Frad}
to the Lorentz force equation for static and spatially uniform
electromagnetic field.

The squared norm (\ref{ABua}) of $O^{(a)}$-projection of particle's
four-velocity decreases exponentially as the time parameter
increases. It describes the decaying circular-like orbit in plane
${\cal M}^{(a)}$. We see that the radiation damping suppresses the
particle's oscillation.

The squared norm (\ref{ABub}) of $O^{(b)}$-projection of particle's
four-velocity tends to $-1$ as the time parameter increases.
Contrary to oscillatory motion, the self force almost does not
influence hyperbolic motion. In specific case of purely electric
field ($\omega_a=0$), one can choose $u^\mu_{(a)}(0)=0$. In this
case, the constant $\alpha=0$ and the factor before the squared
brackets in eq. (\ref{uObL}) becomes $1$. The radiation reaction
force on a charged particle vanishes when the particle accelerates
uniformly \cite{FR60}.

The $O^{(b)}$-projection of particle's 4-velocity does not change
under the influence of the self force even if an external field is
very powerful. The greater is the argument $\lambda_b\tau$ of
hyperbolic functions, the closer are their magnitudes:
$$
\sinh(\lambda_b\tau)\approx\pm\cosh(\lambda_b\tau).
$$
So, if $\lambda_b<0$ (negative charge), then $O^{(b)}$-projection of
four-velocity (\ref{uObL}) approaches the null 4-vector
\begin{eqnarray}
u^\mu_{(b)}(\tau)&\approx&\frac{1}{\sqrt{\beta-\alpha{\rm e}^{-2\omega_0\tau}}}\left[u^\mu_{(b)}(0)\cosh(\lambda_b\tau)-b^{-1}F^\mu{}_\alpha u^\alpha_{(b)}(0)\cosh(\lambda_b\tau)\right]\nonumber\\
&=&CP_{(-b)}u^\alpha(0),\nonumber
\end{eqnarray}
where
$$
C=2\frac{{\rm e}^{\omega_0\tau}}{\sqrt{\beta{\rm e}^{2\omega_0\tau}-\alpha}}\cosh(\lambda_b\tau)\,,
$$
and $P_{(-b)}$ is the projection operator (\ref{Pbmn}) which
produces eigenvector with negative real value $-b$. If $\lambda_b>0$
(positive charge), then $O^{(b)}$-projection of four-velocity
(\ref{uObL}) tends to eigenvector with positive real value
\begin{eqnarray}
u^\mu_{(b)}(\tau)&\approx&\frac{1}{\sqrt{\beta-\alpha{\rm e}^{-2\omega_0\tau}}}\left[u^\mu_{(b)}(0)\cosh(\lambda_b\tau)+b^{-1}F^\mu{}_\alpha u^\alpha_{(b)}(0)\cosh(\lambda_b\tau)\right]\nonumber\\
&=&CP_{(+b)}u^\alpha(0).\nonumber
\end{eqnarray}
When the time increases, the orbit in plane ${\cal M}^{(b)}$
approaches the null ray along one of the ``real'' eigenvectors of
the electromagnetic field tensor. Therefore, the equation for
eigenvalues and eigenvectors of the (external) electromagnetic field
tensor is the ultrarelativistic limit of the LL equation. Moreover,
this is true not only for a constant field, but for a field of
arbitrary configuration \cite[Chap.9]{YT}.

\subsection{World line}

For completeness, we give the expressions for the particle's world line
\begin{eqnarray}
\zeta&:&{\mathbb R}\to{\mathbb M}_{\,4}\nonumber\\
&&\tau\mapsto (z^\alpha(\tau))
\end{eqnarray}
when the initial velocity is $u(0)=u_{(a)}(0)+u_{(b)}(0)$ and the
initial position $z(0)=z_{(a)}(0)+z_{(b)}(0)$. Zeroth component
$z^0(\tau)$ gives the laboratory time $t$.

Integrating Eqs. (\ref{uObL}), we obtain the two different
projections of the position 4-vector. For future convenience, we
change the variable $x=\omega_0\tau$ and introduce new parameter
$\lambda=\sqrt{\alpha/\beta}<1$
\begin{eqnarray}
z^\mu_{(a)}(\tau)&=&z^\mu_{(a)}(0)+\frac{1}{\omega_0\sqrt{\alpha}}\left[u^\mu_{(a)}(0)\int{\rm d}x\frac{\lambda{\rm e}^{-x}}{\sqrt{1-(\lambda{\rm e}^{-x})^2}}\cos(\omega_a'x)\right.\nonumber\\
&+&\left.a^{-1}F^\mu{}_\alpha u^\alpha_{(a)}(0)\int{\rm d}x\frac{\lambda{\rm e}^{-x}}{\sqrt{1-(\lambda{\rm e}^{-x})^2}}\sin(\omega_a'x)\right],\label{za}\\
z^\mu_{(b)}(\tau)&=&z^\mu_{(b)}(0)+\frac{1}{\omega_0\sqrt{\beta}}\left[u^\mu_{(b)}(0)\int{\rm d}x\frac{1}{\sqrt{1-(\lambda{\rm e}^{-x})^2}}\cosh(\lambda_b'x)\right.\nonumber\\
&+&\left.b^{-1}F^\mu{}_\alpha u^\alpha_{(b)}(0)\int{\rm d}x\frac{1}{\sqrt{1-(\lambda{\rm e}^{-x})^2}}\sinh(\lambda_b'x)\right].\label{zb}
\end{eqnarray}

To derive the ``a''-projection, we express the time-dependent
amplitude under the integral signs in Eq. (\ref{za}) as
hypergeometric series \cite[Eq.(15.1.8)]{ASBook}
\begin{equation}
\frac{\lambda{\rm e}^{-x}}{\sqrt{1-(\lambda{\rm e}^{-x})^2}}=\sum_{n=0}^\infty
\frac{\Gamma(\frac12+n)}{\Gamma(\frac12)n!}(\lambda{\rm e}^{-x})^{2n+1}.\label{suma}
\end{equation}
To perform calculations, we use the formulae
\begin{eqnarray}
\int_0^x{\rm d}t\,{\rm e}^{qt}\cos(\omega t)
&=&\frac{{\rm e}^{qx}\sin(\omega x+\phi)-\sin\phi}{\sqrt{\omega^2+q^2}},\nonumber\\
\int_0^x{\rm d}t\,{\rm e}^{qt}\sin(\omega t)
&=&-\frac{{\rm e}^{qx}\cos(\omega x+\phi)-\cos\phi}{\sqrt{\omega^2+q^2}},\nonumber
\end{eqnarray}
where phase $\phi$ is given by
$$
\cos\phi=\frac{\omega}{\sqrt{\omega^2+q^2}},\qquad \sin\phi=\frac{q}{\sqrt{\omega^2+q^2}}.
$$
Inserting these into eq.(\ref{za}), we derive the orbit in the plane ${\cal M}^{(a)}$
\begin{equation}\label{za-mu}
z^\mu_{(a)}(\tau)=z^\mu_{(a)}(0)+\frac{1}{\sqrt{\alpha}}\left[u^\mu_{(a)}(0)I_s-a^{-1}F^\mu{}_\alpha u^\alpha_{(a)}(0)I_c\right],
\end{equation}
where symbols $I_s$ and $I_c$ denote the series
\begin{eqnarray}
I_s&=&\sum_{n=0}^\infty a_n\left[{\rm e}^{-(2n+1)\omega_0\tau}\sin(\omega_a\tau+\phi_n)-\sin\phi_n\right],\nonumber\\
I_c&=&\sum_{n=0}^\infty a_n\left[{\rm e}^{-(2n+1)\omega_0\tau}\cos(\omega_a\tau+\phi_n)-\cos\phi_n\right],\nonumber
\end{eqnarray}
with coefficients
\begin{equation}\label{a-n}
a_n=\frac{\Gamma(\frac12+n)}{\Gamma(\frac12)n!}\frac{\lambda^{2n+1}}{\sqrt{\omega_a^2+[(2n+1)\omega_0]^2}}\,.
\end{equation}
Thus, the ``a''-projection  behaves like ensemble of damped harmonic
oscillators with common frequency $\omega_a$ and different phases
\begin{equation}\label{phin}
\cos\phi_n=\frac{\omega_a}{\sqrt{\omega_a^2+[(2n+1)\omega_0]^2}}\,,\qquad
\sin\phi_n=-\frac{(2n+1)\omega_0}{\sqrt{\omega_a^2+[(2n+1)\omega_0]^2}}\,.
\end{equation}{
In the plane ${\cal M}^{(a)}$, the charged particle spirals inward,
i.e., moves in a continuous curve that gets nearer to central point.

Now, integrating equation (\ref{zb}) we obtain the ``b''-projection
of position 4-vector. We expand the time-dependent amplitude as
hypergeometric series (\ref{suma}) and apply the formulae
\begin{eqnarray}
\int_0^x{\rm d}t\,{\rm e}^{qt}\cosh(\lambda t)&=&\frac{{\rm e}^{qx}\sinh(\lambda x-\psi)+\sinh\psi}{\sqrt{\lambda^2-q^2}},\nonumber \\
\int_0^x{\rm d}t\,{\rm e}^{qt}\sinh(\lambda t)&=&\frac{{\rm e}^{qx}\cosh(\lambda x-\psi)-\cosh\psi}{\sqrt{\lambda^2-q^2}},\nonumber
\end{eqnarray}
where phase $\psi$ is given by
$$
\cosh\psi=\frac{\lambda}{\sqrt{\lambda^2-q^2}},\qquad \sinh\psi=\frac{q}{\sqrt{\lambda^2-q^2}}.
$$
The result is
\begin{equation}\label{zb-mu}
z^\mu_{(b)}(\tau)=z^\mu_{(b)}(0)+\frac{1}{\sqrt{\beta}}\left[u^\mu_{(b)}(0)J_s+b^{-1}F^\mu{}_\alpha u^\alpha_{(b)}(0)J_c\right],
\end{equation}
where $J_s$ and $J_c$ are time-dependent series
\begin{eqnarray}
J_s&=&\sum_{n=0}^\infty b_n\left[{\rm e}^{-2n\omega_0\tau}\sinh(\lambda_b\tau-\psi_n)+\sinh\psi_n\right],\nonumber\\
J_c&=&\sum_{n=0}^\infty b_n\left[{\rm e}^{-2n\omega_0\tau}\cosh(\lambda_b\tau-\psi_n)-\cosh\psi_n\right],\nonumber
\end{eqnarray}
with coefficients
\begin{equation}\label{b-n}
b_n=\frac{\Gamma(\frac12+n)}{\Gamma(\frac12)n!}\frac{\lambda^{2n}}{\sqrt{\lambda_b^2-[2n\omega_0]^2}}\,.
\end{equation}
$n$-th shift of the hyperbolic phase is given by
$$
\cosh\psi_n=\frac{\lambda_b}{\sqrt{\lambda_b^2-[2n\omega_0]^2}},\qquad \sinh\psi_n=-\frac{2n\omega_0}{\sqrt{\lambda_b^2-[2n\omega_0]^2}}.
$$

A charged particle acted upon by a static spatially uniform electric
field moves along hyperbola characterized by an inverse time
constant $e|\mathbb{E}|/m$; a hyperbolic phase is determined by an
initial velocity. Therefore, the ``b''-projection looks like
ensemble of identical charges with different (discrete) initial
velocities moving in constant electric field. In the plane ${\cal
M}^{(b)}$, the charged particle exhibits an exponential proper time
behavior: it asymptotically approaches the null ray aligned along
one of two ``real'' eigenvectors of the electromagnetic field
tensor.

Finally, using the relation $z(\tau)=z_{(a)}(\tau)+z_{(b)}(\tau)$,
the two preceding projections may be combined to give the total
position 4-vector. To simplify future calculations, we present the
product relations \cite[Eqs.(2.5),(2.6)]{Frad}
\begin{eqnarray}
F\cdot F&=&(b^2-a^2)I+\!\!\!\phantom{1}^*\!F\cdot\!\!\!\phantom{1}^*\!F,\label{FF}\\
F\cdot\!\!\!\phantom{1}^*\!F&=&\!\!\!\phantom{1}^*\!F\cdot F\nonumber\\
&=&-\kappa abI,\label{FdF}
\end{eqnarray}
where $\kappa={\rm sign}{\cal P}$ and ``dot'' denotes the matrix
product, e.g., $F^\mu{}_\alpha F^\alpha{}_\beta $. After some
algebra one can also deduce \cite[eqs.(2.7)]{Frad}
\begin{equation}
F\cdot F\cdot F=(b^2-a^2)F-\kappa ab\!\!\!\phantom{1}^*\!F.
\end{equation}
The Hamilton-Cayley equation (\ref{chrc}) is not necessary in this
calculations because the fourth power of $F$ does not appear in Eqs.
(\ref{za-mu}) and (\ref{zb-mu}).

\subsection{Radiation emission}
The rate of energy-momentum loss due to radiation is given by the
Abraham-Heaviside formula
\begin{equation}\label{AHf}
\frac{{\rm d}P^\mu}{{\rm d}\tau}=-\frac{2e^2}{3}(a\cdot a)u^\mu.
\end{equation}
Particle's 4-velocity which satisfies the Lorentz-Abraham-Dirac
equation (\ref{LAD}) and its proper-time derivative as
4-acceleration should be substituted. According to Teitelboim's
careful examination \cite{Teit}, this is the energy-momentum carried
by electromagnetic waves that detach themselves from the charge and
lead an independent existence.

The zeroth component of 4-vector (\ref{AHf}) is the rate of emitted
energy which can be detected dy distant devices. According to
\cite[eq. (35)]{HL10}, the rate with respect to laboratory time is
of true physical sense:
\begin{equation}\label{AH-f}
\frac{{\rm d}P^0}{{\rm d}t}=-\frac{2e^2}{3}(a\cdot a).
\end{equation}
To calculate it, we use the solution (\ref{uObL}) to the
Landau-Lifshitz approximation of the Lorentz-Abraham-Dirac equation.
The square of its proper-time derivative is as follows:
\begin{eqnarray}\label{a-a}
(a\cdot a)&=&a_{(a)}^2+ a_{(b)}^2\nonumber\\
&=&\lambda_b^2+\frac{(\lambda{\rm e}^{-\omega_0\tau})^2}{1-(\lambda{\rm e}^{-\omega_0\tau})^2}\left(\lambda_b^2+\omega_a^2\right)+\frac{(\lambda{\rm e}^{-\omega_0\tau})^2}{\left[1-(\lambda{\rm e}^{-\omega_0\tau})^2\right]^2}\,\omega_0^2\,,
\end{eqnarray}
where $\lambda=\sqrt{\alpha/\beta}<1$.
Finally, with a degree of accuracy sufficient for our purposes
\begin{equation}\label{rate}
\frac{{\rm d}P^0}{{\rm d}t}=-\frac{2e^2}{3}\lambda_b^2-m\omega_0\frac{(\lambda{\rm e}^{-\omega_0\tau})^2}{1-(\lambda{\rm e}^{-\omega_0\tau})^2},
\end{equation}
where $\omega_0$ is a constant defined by Eq. (\ref{Om_0}).

We see that the intensity of radiation strongly depends on time.
(Strictly speaking, the proper time in the right-hand side of Eq.
(\ref{rate}) should be substituted for the laboratory time. Since
$t=z^0(\tau)$, the function $\tau(t)$ is inverse to the zeroth
coordinate function $z^0(\tau)$ which is given by eqs. (\ref{za-mu})
and (\ref{zb-mu}). It is of great importance that both the proper
time and laboratory time are monotonic increasing quantities.)
Initially, the synchrotron radiation and the longitudinal radiation
present in almost equal quantities. But the radiation damping
suppresses the ``magnetic'' rotation, so that the longitudinal
radiation (this caused by uniformly accelerated charge) only
survives. This is in line the main result by Fulton and Rohrlich
\cite[Eq.(3.8)]{FR60}. So far as the synchrotron radiation is
concerned, Shen \cite{Shen} also states that it decreases with time
as well as the influence of the radiation damping force.

\section{Applications}\label{Applicat}
\setcounter{equation}{0}

We now discuss our main examples of dynamics of a radiating charge
traveling in constant electromagnetic fields of magnetic type and of
electric type as well as in the crossed field. We demonstrate how
the projection technique works and compare the results with these
known in literature \cite{Ares,Her73,Her74,Her70,Spohn}.

Assume that in the laboratory frame the state of the electromagnetic
field at point $x\in{\mathbb M}_{\,4}$ is specified by three
components $(E^1,E^2,E^3)$ of electric field $\mathbf E$ and three
components $(B^1,B^2,B^3)$ of magnetic field $\mathbf B$. If the
inertial frame moves with a velocity $\mathbf v$ with respect to
laboratory one, then the new strengths become
\begin{eqnarray}\label{Eprime}
{\mathbf E}'&=&u^0{\mathbf E}-\frac{({\mathbf u}{\mathbf E}){\mathbf u}}{u^0+1}+[{\mathbf u}\times {\mathbf B}],\\
{\mathbf B}'&=&u^0{\mathbf B}-\frac{({\mathbf u}{\mathbf B}){\mathbf u}}{u^0+1}-[{\mathbf u}\times {\mathbf E}].\label{Bprime}
\end{eqnarray}
$u^0$ is zeroth component of normalized four-velocity
$u=(\gamma,\gamma v^1,\gamma v^2,\gamma v^3)$ which relates the
inertial frames. Under a spatial rotation, the matrices (\ref{Far})
and (\ref{Fst}) transform in such a way that electric field $\mathbf
E$ and magnetic field $\mathbf B$ transform as three-vectors.

\subsection{Motion in a field of magnetic type}

If invariants ${\cal P}=0$ and ${\cal S}>0$, then in all reference
frames the magnitude of magnetic field is greater than that of
electric field.  The Lorentz transformation which is determined by
4-velocity
\begin{equation}\label{uB}
u^0=\frac{|\mathbf B|}{\sqrt{S}},\qquad
{\mathbf u}=\frac{[{\mathbf E}\times{\mathbf B}\,]}{|{\mathbf B}|\sqrt{\cal S}}
\end{equation}
emphasizes the privileged inertial frame where the electric field
${\mathbf E}'=0$. The primed magnetic field (\ref{Bprime})
$$
{\mathbf B}'=\sqrt{\cal S}{\mathbf n}_B
$$
is the only one which ``survives'' in the privileged inertial frame.
Here ${\mathbf n}_B={\mathbf B}/|{\mathbf B}|$ is the unit vector in
$B$-direction. If we align the $z$-axis along ${\mathbf B}'$, the
electromagnetic field tensor (\ref{FldT}) takes the form
\begin{equation}\label{FlTB}
(F^\alpha{}_\nu)=\left(
\begin{array}{cccc}
0&0&0&0\\
0&0&a&0\\
0&-a&0&0\\
0&0&0&0\\
\end{array}
\right),
\end{equation}
where $a=\sqrt{\cal S}$ is the magnitude of imaginary eigenvalue.
The set of eigenvalues consists of doubly degenerate zero and pair
of imaginary ones, $+{\rm i}a$ and $-{\rm i}a$ (see Sec.~\ref{Inv}).

Taking into account that $b=0$, we derive the projection operators
\begin{eqnarray}
O^{(a)}&=&-\frac{F^2}{a^2}\nonumber\\
&=&\left(
\begin{array}{cccc}
0&0&0&0\\
0&1&0&0\\
0&0&1&0\\
0&0&0&0\\
\end{array}
\right),\label{Oa-m}
\end{eqnarray}
and
\begin{eqnarray}
O^{(b)}&=&\frac{F^2+a^2I}{a^2}\nonumber\\
&=&\left(
\begin{array}{cccc}
1&0&0&0\\
0&0&0&0\\
0&0&0&0\\
0&0&0&1\\
\end{array}
\right).\label{Ob-m}
\end{eqnarray}
The ``a projection'' of particle's 4-velocity has coordinates
$(0,u^1,u^2,0)$ while $u_{(b)}=(u^0,0,0,u^3)$. Operators
(\ref{Oa-m}) and (\ref{Ob-m}) split the tangent space ${\rm
T}_x{\mathbb M}_{\,4}$ into two mutually orthogonal subspaces ${\cal
M}_x^{(a)}=\{v\in{\rm T}_x{\mathbb M}_{\,4}:v^0=0,v^3=0\}$ and
${\cal M}_x^{(b)}=\{v\in{\rm T}_x{\mathbb M}_{\,4}:v^1=0,v^2=0\}$.
Traveling from point to point, we cover the affine space ${\mathbb
M}_{\,4}$ by coordinate grid composed from mutually orthogonal
planes of two dimensions. The projection operators
\begin{equation}\label{Ppa}
P_{(+a)}=\frac12\left(
\begin{array}{cccc}
0&0&0&0\\
0&1&-{\rm i}&0\\
0&{\rm i}&1&0\\
0&0&0&0\\
\end{array}
\right)
\end{equation}
and
\begin{equation}\label{Pma}
P_{(-a)}=\frac12\left(
\begin{array}{cccc}
0&0&0&0\\
0&1&{\rm i}&0\\
0&-{\rm i}&1&0\\
0&0&0&0\\
\end{array}
\right)
\end{equation}
produce the eigenvectors of the field tensor (\ref{FlTB}) which
correspond to eigenvalues $+{\rm i}a$ and $-{\rm i}a$, respectively.

The contractions of the field tensor (\ref{FlTB}) and velocity
projections $u_{(a)}$ and $u_{(b)}$ are as follows:
$$
a^{-1}F^\mu{}_\alpha u^\alpha_{(a)}=(0,u^2,-u^1,0),\qquad F^\mu{}_\alpha u^\alpha_{(b)}=(0,0,0,0).
$$
The solution (\ref{uObL}) of the Herrera equation simplifies in this
specific case
\begin{eqnarray}\label{u-mgn}
u^0(\tau)&=&\frac{\sqrt{\beta}\cosh\chi_0}{\sqrt{\beta-\alpha{\rm e}^{-2\omega_0\tau}}},\nonumber\\
u^1(\tau)&=&\frac{\sqrt{\alpha}\cos(\omega_a\tau+\varphi_0)}{\sqrt{\beta{\rm e}^{2\omega_0\tau}-\alpha}},\nonumber\\
u^2(\tau)&=&-\frac{\sqrt{\alpha}\sin(\omega_a\tau+\varphi_0)}{\sqrt{\beta{\rm e}^{2\omega_0\tau}-\alpha}},\nonumber\\
u^3(\tau)&=&\frac{\sqrt{\beta}\sinh\chi_0}{\sqrt{\beta{-\alpha\rm e}^{-2\omega_0\tau}}},
\end{eqnarray}
where both the initial hyperbolic phase
\begin{equation}\label{chi0}
\cosh\chi_0=\frac{u^0(0)}{\sqrt{\beta}},\qquad \sinh\chi_0=\frac{u^3(0)}{\sqrt{\beta}},
\end{equation}
and the initial trigonometric phase
\begin{equation}\label{vrp0}
\cos\varphi_0=\frac{u^1(0)}{\sqrt{\alpha}},\qquad \sin\varphi_0=-\frac{u^2(0)}{\sqrt{\alpha}},
\end{equation}
are defined by the components of the initial 4-velocity. Denoting
$\beta=\gamma_0^2$ and $\alpha=\gamma_0^2-1$, we are sure that the
result is in accord with that obtained in \cite[Eqs.
(19),(20)]{Ares}.

In series of papers \cite{Her70,Her73,Her74}, Herrera studied the
behavior of radiating charge acted upon by a uniform magnetic field
${\mathbf H}=(0,0,H)$. The Lorentz-Abraham-Dirac equation
(\ref{LAD}) is chosen as the equation of motion of this charge. The
solution is presented as series in powers of dimensionless small
parameter
$$
\lambda=\frac{2e^2}{3mc^3}\frac{e}{m}\frac{H}{c}.
$$
(The units are Gaussian.) The equation (\ref{LLeq}) is the first
order approximation in Herrera's approach. (Zeroth approximation
gives the Lorentz force equation with no radiation.) The effect of
radiation damping to the first power of the expansion parameter is
illustrated by Eqs. (22) and (23) in \cite{Her73} which are just the
expressions (\ref{u-mgn}) for the components of particle's
4-velocity.

In Ref. \cite[eq.(5)]{Her74}, the relation between the proper time
$\tau$ and the laboratory time $t$ is found. In the first order
degree of accuracy, Herrera's function $Z(t)$ is proportional to
$t$. The relation is presented in the form of the Lorentz factor
\begin{equation}
\gamma(t)=\cosh\chi_0\frac{\sqrt{\beta}+1+(\sqrt{\beta}-1){\rm e}^{-2\omega_0't}}{\sqrt{\beta}+1-(\sqrt{\beta}-1){\rm e}^{-2\omega_0't}},
\end{equation}
where
$$
\omega_0'=\frac{\omega_0}{\cosh\chi_0},
$$
(see also \cite[eq. (9.20)]{Spohn}). Inverting the relation
$u^0(\tau)=\gamma(t)$ where the zeroth component is written in the
first line of eqs. (\ref{u-mgn}), we obtain
\begin{equation}\label{tautB}
\tau(t)=\frac{1}{\omega_0}{\rm ln}\left[\cosh(\omega_0't)+\beta^{-1/2}\sinh(\omega_0't)\right].
\end{equation}
Substituting this into Eq. (\ref{rtt}), we obtain the rate of
radiated energy with respect to the laboratory time
\begin{equation}\label{rtt}
\frac{{\rm d}P^0}{{\rm d}t}=-\omega_0\frac{\alpha}{\left[\sqrt{\beta}\cosh(\omega_0't)+\sinh(\omega_0't)\right]^2}.
\end{equation}
The intensity of radiation decreases with time exponentially.

Our next task is to derive the coordinate functions. Having
integrated  the zeroth component of 4-velocity, we obtain the zeroth
coordinate
\begin{equation}
z^0(\tau)=z^0(0)+\frac{u^0(0)}{\sqrt{\beta}\omega_0}{\rm ln}
\frac{\sqrt{\beta}{\rm e}^{\omega_0\tau}+\sqrt{\beta{\rm
e}^{2\omega_0\tau}-\alpha}}{\sqrt{\beta}+1},
\end{equation}
which gives the laboratory time $t=z^0(\tau)$. We choose the
constant of integration such that the laboratory watch and the
particle's watch show exactly the same time at the beginning of
particle's motion
\begin{equation}
t=\frac{1}{\omega_0'}{\rm ln}
\frac{\sqrt{\beta}{\rm e}^{\omega_0\tau}+\sqrt{\beta{\rm
e}^{2\omega_0\tau}-\alpha}}{\sqrt{\beta}+1}.
\end{equation}
The longitudinal motion has the form $z^3(t)=z^3(0)+\tanh\chi_0 t$
in the laboratory time parametrization.

In the plane which is orthogonal to the magnetic field vector, the
charge moves on the decaying spiral
\begin{eqnarray}
z^1(\tau)&=&z^1(0)+\sum_{n=0}^\infty a_n\left[{\rm e}^{-(2n+1)\omega_0\tau}\sin(\omega_a\tau+\varphi_0+\phi_n)-\sin(\varphi_0+\phi_n)\right],\nonumber\\
z^2(\tau)&=&z^2(0)+\sum_{n=0}^\infty a_n\left[{\rm e}^{-(2n+1)\omega_0\tau}\cos(\omega_a\tau+\varphi_0+\phi_n)-\cos(\varphi_0+\phi_n)\right],\nonumber\\
\label{z12}
\end{eqnarray}
where $\phi_n$ is given by Eqs. (\ref{phin}) and the phase's shift
$\varphi_0$ is defined by Eq. (\ref{vrp0}).

Let us assess how quickly the charge gets nearer to the central
point with coordinates
\begin{eqnarray}
z^1_c&=&z^1(0)-\sum_{n=0}^\infty a_n\sin(\varphi_0+\phi_n),\nonumber\\
z^2_c&=&z^2(0)-\sum_{n=0}^\infty a_n\cos(\varphi_0+\phi_n).\nonumber
\end{eqnarray}
To sum up the series in Eqs. (\ref{z12}), we neglect the shifts
$\phi_n$ and the second term under the square root in the
denominator of $a_n$. In this approximation, the $n$-dependent terms
constitute the hypergeometric series. After some algebra, we obtain
$$
z^1(\tau)=z^1(0)+\frac{\sqrt{\alpha}\sin(\omega_a\tau+\varphi_0)}{\omega_a\sqrt{\beta{\rm e}^{2\omega_0\tau}-\alpha}},\qquad
z^2(\tau)=z^2(0)+\frac{\sqrt{\alpha}\cos(\omega_a\tau+\varphi_0)}{\omega_a\sqrt{\beta{\rm e}^{2\omega_0\tau}-\alpha}}.
$$
Substituting the laboratory time for the proper time in the
expression for the decreasing radius
$r=\sqrt{[z^1(\tau)-z^1(0)]^2+[z^2(\tau)-z^2(0)]^2}$ of the spiral,
we obtain the formula
\begin{eqnarray}
r(t)&=&\frac{\sqrt{\alpha}}{\omega_a}\frac{1}{\sqrt{\beta{\rm e}^{2\omega_0\tau}-\alpha}}\nonumber\\
&=&\frac{\sqrt{(v^1_0)^2+(v^2_0)^2}}{\gamma_0\omega_a\left[\cosh(\omega_0't)+\sqrt{\beta}\sinh(\omega_0't)\right]}
\end{eqnarray}
which was first obtained by Spohn \cite[eqs. (9.24)-(9.25)]{Spohn}.
In fact, the radius decreases even more rapidly because we
substitute the larger amplitude $\omega_a^{-1}$ for the smaller one
$\left\{\omega_a^2+[(2n+1)\omega_0]^2\right\}^{-1/2}$ in the each coefficient
(\ref{a-n}) of series (\ref{z12}).

\subsection{Motion in a field of electric type}

If invariants ${\cal P}= 0$ and ${\cal S}< 0$, then in all reference
frames the magnitude of electric field is greater than that of
magnetic field. The Lorentz transformation which is determined by
4-velocity
\begin{equation}\label{uE}
u^0=\frac{|{\mathbf E}|}{\sqrt{-{\cal S}}},\qquad
{\mathbf u}=\frac{[{\mathbf E}\times{\mathbf B}\,]}{|{\mathbf E}|\sqrt{-{\cal S}}}
\end{equation}
emphasizes the privileged inertial frame where the magnetic field
${\mathbf B}'=0$. The primed electric field (\ref{Eprime})
$$
{\mathbf E}'=\sqrt{-{\cal S}}{\mathbf n}_E
$$
is the only one which ``survives'' in the privileged inertial frame.
Here, ${\mathbf n}_E={\mathbf E}/|{\mathbf E}|$ is the unit vector in
$E$-direction. If we align the $z$-axis along ${\mathbf E}'$, the
electromagnetic field tensor (\ref{FldT}) takes the form
\begin{equation}\label{FlTE}
(F^\alpha{}_\nu)=\left(
\begin{array}{cccc}
0&0&0&b\\
0&0&0&0\\
0&0&0&0\\
b&0&0&0\\
\end{array}
\right),
\end{equation}
where $b=\sqrt{-{\cal S}}$ is the magnitude of real eigenvalue. The
set of eigenvalues consists of doubly degenerate zero and pair of
real ones, $+b$ and $-b$ (see Sec. \ref{Inv}).

Fradkin's operators $O^{(a)}$ and $O^{(b)}$ have the form
(\ref{Oa-m}) and (\ref{Ob-m}), respectively. They produce the planes
${\cal M}^{(a)}$ and ${\cal M}^{(b)}$ which has been described in
preceding Paragraph. The projection operators
\begin{equation}\label{Ppb}
P_{(+b)}=\frac12\left(
\begin{array}{cccc}
1&0&0&1\\
0&0&0&0\\
0&0&0&0\\
1&0&0&1\\
\end{array}
\right)
\end{equation}
and
\begin{equation}\label{Pmb}
P_{(-b)}=\frac12\left(
\begin{array}{cccc}
1&0&0&-1\\
0&0&0&0\\
0&0&0&0\\
-1&0&0&1\\
\end{array}
\right)
\end{equation}
produce the eigenvectors of the field tensor (\ref{FlTE}) which
correspond to eigenvalues $+b$ and $-b$, respectively.

The contractions of this tensor and velocity projections
$u_{(a)}=(0,u^1,u^2,0)$ and $u_{(b)}=(u^0,0,0,u^3)$ are as follows:
$$
F^\mu{}_\alpha u^\alpha_{(a)}=(0,0,0,0),\qquad b^{-1}F^\mu{}_\alpha u^\alpha_{(b)}=(u^3,0,0,u^0).
$$
Substituting this into Eqs. (\ref{uObL}), we arrive at
\begin{eqnarray}
u^0(\tau)&=&\frac{\sqrt{\beta}\cosh(\lambda_b\tau+\chi_0)}{\sqrt{\beta-\alpha{\rm e}^{-2\omega_0\tau}}},\label{u0E}\\
u^1(\tau)&=&\frac{\sqrt{\alpha}\cos\varphi_0}{\sqrt{\beta{\rm e}^{2\omega_0\tau}-\alpha}},\label{u1E}\\
u^2(\tau)&=&-\frac{\sqrt{\alpha}\sin\varphi_0}{\sqrt{\beta{\rm e}^{2\omega_0\tau}-\alpha}},\label{u2E}\\
u^3(\tau)&=&\frac{\sqrt{\beta}\sinh(\lambda_b\tau+\chi_0)}{\sqrt{\beta-\alpha{\rm e}^{-2\omega_0\tau}}}.\label{u0-3E}
\end{eqnarray}
If the initial values $u^1(0)=0$ and $u^2(0)=0$ in the privileged
reference frame determined by Eqs. (\ref{uE}), then the constants
$\alpha=0$ and $\beta=1$. The simplified version of the expressions
(\ref{u0E})--(\ref{u0-3E})
\begin{eqnarray}
u^0(\tau)&=&\cosh(\lambda_b\tau+\chi_0),\nonumber\\
u^1(\tau)&=&0,\nonumber\\
u^2(\tau)&=&0,\nonumber\\
u^3(\tau)&=&\sinh(\lambda_b\tau+\chi_0),\label{u03E}
\end{eqnarray}
has been discussed in detail by Fulton and Rohrlich \cite{FR60}. It
is worth noting that the velocity components define the exact
solution to the Lorentz-Abraham-Dirac equation. In the absence of
any transverse disturbances, the constant electric field supplies
the amount of energy and momentum exactly equal to the energy and
momentum losses due to radiation.

If the transverse components of initial 4-velocity do not vanish in
the privileged inertial frame, the position functions $z^0(\tau)$
and $z^3(\tau)$ are given by the series
\begin{eqnarray}
z^0(\tau)&=&z^0(0)+\sum_{n=0}^\infty b_n\left[{\rm e}^{-2n\omega_0\tau}\sinh(\lambda_b\tau-\psi_n+\chi_0)-\sinh(-\psi_n+\chi_0)\right],\nonumber\\
z^3(\tau)&=&z^3(0)+\sum_{n=0}^\infty b_n\left[{\rm e}^{-2n\omega_0\tau}\cosh(\lambda_b\tau-\psi_n+\chi_0)-\cosh(-\psi_n+\chi_0)\right],\nonumber\\
\label{z03}
\end{eqnarray}
where the hyperbolic phase shift $\chi_0$ is determined by Eq.
(\ref{chi0}) and coefficients $b_n$ are defined by Eq. (\ref{b-n}).
The orbit in the transverse plane ${\cal M}^{(a)}$ is given by the
coordinate functions
\begin{eqnarray}
z^1(\tau)&=&z^1(0)+\frac{\cos\varphi_0}{\omega_0}\left[\arccos(\lambda{\rm e}^{-\omega_0\tau})-\arccos\lambda\right],\nonumber\\
z^2(\tau)&=&z^2(0)-\frac{\sin\varphi_0}{\omega_0}\left[\arccos(\lambda{\rm e}^{-\omega_0\tau})-\arccos\lambda\right],
\end{eqnarray}
which we derive by means of integration of the right-hand sides of
Eqs. (\ref{u1E}) and (\ref{u2E}). If the proper time parameter
increases, the transverse orbit tends to the point with coordinates
\begin{eqnarray}
z^1_\infty&=&z^1(0)+\frac{\cos\varphi_0}{\omega_0}\left[\frac{\pi}{2}-\arccos\lambda\right],\nonumber\\
z^2_\infty&=&z^2(0)-\frac{\sin\varphi_0}{\omega_0}\left[\frac{\pi}{2}-\arccos\lambda\right],\nonumber
\end{eqnarray}
at which the components $u^1_\infty=0$ and $u^2_\infty=0$. The orbit
in  the longitudinal plane ${\cal M}^{(b)}$ approaches the
hyperbola.

\subsection{Motion in a crossed field}

If both the invariants ${\cal S}\neq 0$ and ${\cal P}\neq 0$, there
exists the inertial frame in which the electric and magnetic fields
are collinear. In terms of their magnitudes $a=|{\mathbf  B}'|$ and
$b=|{\mathbf  E}'|$, the invariants (\ref{SP}) look as follows:
\begin{eqnarray}
{\cal P}=2\kappa ab,\qquad {\cal S}=a^2-b^2.\nonumber
\end{eqnarray}
The sign parameter $\kappa=+1$ if ${\mathbf  E}'$ and ${\mathbf B}'$
are in the same direction while $\kappa=-1$ if these vectors are
directed oppositely. The real and positive solutions of this system
of two algebraic equations are just the scalars (\ref{aeig}) and
(\ref{beig}).

The Lorentz transformation to privileged inertial frame is
determined by three-velocity
\begin{equation}\label{uEB}
{\mathbf  u}=\frac{[{\mathbf  E}\times{\mathbf  B}\,]}{\sqrt{d(a^2+b^2)}},
\end{equation}
where
\begin{eqnarray}
d&=&{\mathbf  B}^2+b^2\nonumber\\
&=&{\mathbf  E}^2+a^2\nonumber\\
&=&\frac12\left[{\mathbf  B}^2+{\mathbf  E}^2+\sqrt{{\cal S}^2+{\cal P}^2}\right].\nonumber
\end{eqnarray}
Inserting this in Eqs. (\ref{Eprime}) and (\ref{Bprime}), we obtain
the ``primed'' fields
\begin{eqnarray}
{\mathbf  E}'&=&\frac{b^2{\mathbf  E}+\kappa ab{\mathbf  B}}{\sqrt{d(a^2+b^2)}},\nonumber\\
{\mathbf  B}'&=&\frac{a^2{\mathbf  B}+\kappa ab{\mathbf  E}}{\sqrt{d(a^2+b^2)}}.\nonumber
\end{eqnarray}
Recall that ${\mathbf  E}$ and ${\mathbf  B}$ are electric and
magnetic field strengths in a laboratory reference frame. It is easy
to show that $\kappa a{\mathbf  E}'=b{\mathbf  B}'$.

In the reference frame where the ``primed'' electric field is
directed along $Oz$-axis, the field tensor (\ref{FldT}) takes the
form
\begin{equation}\label{FldTEB}
(F^\alpha{}_\nu)=\left(
\begin{array}{cccc}
0&0&0&b\\
0&0&\kappa a&0\\
0&-\kappa a&0&0\\
b&0&0&0\\
\end{array}
\right).
\end{equation}
Fradkin's operators $O^{(a)}$ and $O^{(b)}$ are still given by Eqs.
(\ref{Oa-m}) and (\ref{Ob-m}), respectively. The generalizations of
projection operators (\ref{Ppa}) and (\ref{Pma})
\begin{equation}
P_{(+a)}=\frac12\left(
\begin{array}{cccc}
0&0&0&0\\
0&1&-{\rm i}\kappa&0\\
0&{\rm i}\kappa&1&0\\
0&0&0&0\\
\end{array}
\right)
\end{equation}
and
\begin{equation}
P_{(-a)}=\frac12\left(
\begin{array}{cccc}
0&0&0&0\\
0&1&{\rm i}\kappa&0\\
0&-{\rm i}\kappa&1&0\\
0&0&0&0\\
\end{array}
\right)
\end{equation}
produce the eigenvectors of the field tensor (\ref{FldTEB}) which
correspond to eigenvalues $+{\rm i}a$ and $-{\rm i}a$, respectively.
The operators (\ref{Ppb}) and (\ref{Pmb}) generate ``real''
eigenvectors.

The contractions of tensor (\ref{FldTEB}) and velocity projections
$u_{(a)}=(0,u^1,u^2,0)$ and $u_{(b)}=(u^0,0,0,u^3)$ are as follows:
$$
a^{-1}F^\mu{}_\alpha u^\alpha_{(a)}=(0,\kappa u^2,-\kappa u^1,0),\qquad b^{-1}F^\mu{}_\alpha u^\alpha_{(b)}=(u^3,0,0,u^0).
$$
Hence, the 4-velocity of the charged particle has the components
\begin{eqnarray}
u^0(\tau)&=&\frac{\sqrt{\beta}}{\sqrt{\beta-\alpha{\rm e}^{-2\omega_0\tau}}}\cosh(\lambda_b\tau+\chi_0),\nonumber\\
u^1(\tau)&=&\frac{\sqrt{\alpha}}{\sqrt{\beta{\rm e}^{2\omega_0\tau}-\alpha}}\cos(\omega_a\tau+\kappa\varphi_0),\nonumber\\
u^2(\tau)&=&-\frac{\sqrt{\alpha}}{\sqrt{\beta{\rm e}^{2\omega_0\tau}-\alpha}}\sin(\omega_a\tau+\kappa\varphi_0),\nonumber\\
u^3(\tau)&=&\frac{\sqrt{\beta}}{\sqrt{\beta-\alpha{\rm e}^{-2\omega_0\tau}}}\sinh(\lambda_b\tau+\chi_0),\nonumber\\
\end{eqnarray}
where initial phases $\varphi_0$ and $\chi_0$ are given by Eqs.
(\ref{vrp0}) and (\ref{chi0}), respectively. Corresponding
coordinate functions which determine the orbit in plane ${\cal
M}^{(a)}$ are given by series (\ref{z12}), and the orbit in plane
${\cal M}^{(b)}$ are defined by series (\ref{z03}).

\section{Conclusions}\label{Concl}
The projection technique developed in this paper allows us to
``visualize'' the formal solution (\ref{APM}) of the LL equation for
a radiating charge in a constant electromagnetic field. The method
can be applied to all types of fields, excepting ones of the null
type (see Sec. \ref{Inv}).

The charge's world line is the combination of two orbits which
``live'' in two mutually orthogonal planes of two dimensions. The
eigenvectors of the pair of imaginary eigenvalues, $+{\rm i}a$ and
$-{\rm i}a$, span the plane ${\cal M}^{(a)}$. The orbit in ${\cal
M}^{(a)}$ is a continuous curve that curves around and gets nearer
to the ``end'' point to which the charge approximately approaches.
The orbit in the plane ${\cal M}^{(b)}$ spanned by eigenvectors of
the pair of real eigenvalues, $+b$ and $-b$, is the slightly
modified hyperbola.

The radiation damping suppresses the particle's rotation in ${\cal
M}^{(a)}$ while the radiation reaction force almost does not
influence on the form of orbit in ${\cal M}^{(b)}$. The world line
is the helix spiralled inward, i. e., a continuous curve that curves
and rises around the line aligned with one of ``real'' eigenvectors
of the electromagnetic field tensor. The radius of decaying orbit
shrinks under the emission of synchrotron radiation.

Let us evaluate the rapidity of radiation damping. The radius of the
circular-like orbit decreases with time exponentially. The argument
of the exponential function is proportional to the constant
$\omega_0=\tau_0(\omega_a^2+\lambda_b^2)$ which defines the
intensity of emission of energy. The stronger the electromagnetic
field, the greater is $\omega_0$, and the shorter is the (proper)
time interval $\Delta\tau$ during which the argument
$-\omega_0\Delta\tau$ of exponential function reaches $-1$. (This
interval is necessary to make the magnitude of the ``a projection''
of particle's 4-velocity smaller in ${\rm e}\approx 2.71828$ times.)
So, if an electron moves in Earth's magnetic field ($|{\vec
B}|\approx 3\cdot 10^{-5}T$), $\Delta\tau\approx 181\,years$. In a
magnetic field of typical refrigerator magnet ($|{\vec B}|\approx
5\cdot 10^{-3}T$), the time interval is $\Delta\tau\approx
57\,hours$.

Much more stronger magnetic field $(|{\vec B}|=5\div10 T)$ is used
in the so-called Penning trap \cite{BG86} where a combination of a
strong homogeneous magnetic field and quadrupole electrostatic
potential is applied to a charged particle to confine it during a
very long time. (During the experiments described in Ref.
\cite{GDK85}, a single electron was trapped continuously for more
than 10 months.) So, if an electron moves within a processing
chamber of a Penning trap, then the interval $\Delta\tau\approx
0.143\div 0.051 s$. The precision measurement \cite{GD85} ($B=6 T$)
results that ``The average excitation energy decreases exponentially
as a function of the delay time with a time constant of $0.27\pm
0.04 s.$'' Radiation damping is the dominant mechanism of reducing
of energy of the cyclotron motion of an electron in a Penning trap
\cite{BG86}.

Extremely powerful magnetic field ($|{\vec B}|\approx 10^6\div 10^8
T$) is in a neighborhood of pulsar. (Pulsar is a rapidly rotating
neutron star producing radiation and radio waves in regular
amounts.) In a pulsar's magnetosphere, the oscillations of electrons
(positrons) are suppressed almost instantly: $\Delta\tau\approx 5.16
ps\div 0.52 fs$. The suppression of electron's oscillations caused
by the radiation damping was first used by Rylov in series of papers
\cite{RlJ,Rl8,Rl9} devoted to the study of plasma's behavior in a
pulsar's magnetosphere. The author considers the equation for
eigenvalue and eigenvectors of the electromagnetic field tensor as
ultrarelativistic limit of the Landau-Lifshitz equation.

\section*{Acknowledgments}

The author wishes to thank V. Tretyak and A. Duviryak for many
useful discussions. I am grateful to referee for a helpful reading
of this manuscript.

\end{document}